\documentclass[
12pt,
preprint,preprintnumbers,nofootinbib,
groupedaddress,superscriptaddress,amsmath,amssymb]{revtex4}
\usepackage{graphicx}
\usepackage{dcolumn}
\usepackage{bm}
\usepackage{amssymb}
\usepackage{amsmath}
\usepackage{epsfig}    
\usepackage{color}
\usepackage{slashed}
\usepackage{hhline}

\def\be{\begin{equation}}
\def\ee{\end{equation}}
\newcommand{\bea}{\begin{eqnarray}}
\newcommand{\eea}{\end{eqnarray}}
\newcommand{\nn}{\nonumber}

\numberwithin{equation}{section}

\begin{document}

\title{Scotogenic neutrino mass with large $SU(2)_L$ multiplet fields}
\preprint{KIAS-P19070, APCTP Pre2019 - 027}
\author{Nilanjana Kumar}
\email{nilanjana.kumar@gmail.com}
\affiliation{Department of Physics and Astrophysics, University of Delhi, Delhi 110007, India}

\author{Takaaki Nomura}
\email{nomura@kias.re.kr}
\affiliation{School of Physics, KIAS, Seoul 02455, Korea}

\author{Hiroshi Okada}
\email{hiroshi.okada@apctp.org}
\affiliation{Asia Pacific Center for Theoretical Physics (APCTP) - Headquarters San 31, Hyoja-dong,
Nam-gu, Pohang 790-784, Korea}
\affiliation{Department of Physics, Pohang University of Science and Technology, Pohang 37673, Republic of Korea}

\date{\today}

\begin{abstract}
 { We construct a scotogenic neutrino mass model introducing large $SU(2)_L$ multiplet fields without adding an extra  symmetry. We have introduced extra scalar fields such as a septet, quintet and quartet where we make the vacuum expectation value of quartet scalar to be zero while septet and quintet develop non-zero ones.  Then the neutrino mass is generated at one-loop level by introducing quintet fermion. We analyze the neutrino mass matrix taking constraints from lepton flavor violation into account and discuss collider physics regarding charged fermions from large multiplet fields. We have analysed the production and the decays of the quintet fermions, as well as the discovery reach at 14 TeV and 27 TeV LHC.}
\end{abstract} 
\maketitle
\newpage

\section{Introduction}
 The standard model (SM) fields are either $SU(2)_L$ singlet or doublet although there is no restriction for the existence of a larger multiplet. 
 In fact we can introduce larger $SU(2)_L$ multiplet fields as exotic field contents which would work to explain the  mystryes in the SM such as non-zero neutrino masses and dark matter, and give rich phenomenologies~\cite{Nomura:2017abu, Cirelli:2005uq,Hambye:2009pw,Sierra:2016qfa,Alvarado:2014jva, Geng:2014oea, Harris:2017ecz, Sierra:2016rcz,Nomura:2016jnl,Nomura:2018ktz,Nomura:2018ibs,Nomura:2018lsx,Nomura:2018cle,Nomura:2018cfu,delAguila:2013yaa,delAguila:2013mia,Chala:2018ari,Aranda:2015xoa}.
For example, models with a septet scalar with hypercharge $Y=2$ have been discussed in Refs.~\cite{Alvarado:2014jva, Geng:2014oea,Aranda:2015xoa, Harris:2017ecz, Sierra:2016rcz} in which $\rho$-parameter is preserved to be $1$ at tree level and Higgs phenomenologies are addressed.
On the other hand, models in Ref.~\cite{Nomura:2016jnl} and \cite{Nomura:2017abu} have also realized the neutrino masses with septet scalar and some other multiplets at tree level and combination of tree and one-loop levels respectively.
In such models, tiny neutrino mass can be partially explained by small vacuum expectation value (VEV) of large multiplet scalar fields, where smallness of these VEVs is also required by the $\rho$-parameter.
Furthermore, large multiplets provide several multiply charged particles which can be produced at collider experiments, which gives interesting phenomenology.

In this paper, we extend the model in Ref.~\cite{Nomura:2017abu} by introducing a quintet scalar field with non-zero VEV.
As a result, tree level neutrino mass can be forbidden and by making quartet scalar an inert scalar, 
neutrino mass is generated at one-loop level. We thus obtain tiny neutrino mass more naturally.
Then the neutrino mass matrix is analyzed taking constants from lepton flavor violation (LFV) into account, and we also estimate muon anomalous magnetic moment and muon $g-2$. Then the collider analysis is done for 
parameter sets satisfying all the constraints.  
In addition, collider phenomenology of exotic particles is different from the previous model since quintet scalar is inert in this model while it develops a VEV in previous one.
We thus discuss exotic particle production at the large hadron collider (LHC) and show signature of our model taking  various decay chains into account. We also project the discovery significance for channels involving 4 b-jets and 
1/2 leptons as a function of the luminosity.


This paper is organized as follows.
In Sec.~II, we introduce our model, derive some formula for active neutrino mass matrix, and show the typical order of Yukawa couplings and related masses.
In Sec.~III, we discuss neutrino mass matrix carrying out numerical analysis and implications to physics at the LHC focusing on the pair production of doubly charged fermion in the multiplet.
We discuss and conclude in Sec.~IV.

\section{ Model setup}
 \begin{widetext}
\begin{center} 
\begin{table}
\begin{tabular}{|c||c|c|c||c|c|c|c|c|}\hline\hline  
&\multicolumn{3}{c||}{Lepton Fields} & \multicolumn{4}{c|}{Scalar Fields} \\\hline
& ~$L_L$~ & ~$e_R^{}$~& ~$\Sigma_R$ ~ & ~$H$ ~ & ~$\Phi_7$~  & ~$\Phi_5$~   & ~$\Phi_4$ \\\hline 
$SU(2)_L$ & $\bm{2}$  & $\bm{1}$  & $\bm{5}$ & $\bm{2}$ & $\bm{7}$ & $\bm{5}$ & $\bm{4}$  \\\hline 
$U(1)_Y$ & $-\frac12$ & $-1$  & $0$  & $\frac12$ & ${1}$ & $0$  & $\frac12$   \\\hline
\end{tabular}
\caption{Contents of fermion and scalar fields
and their charge assignments under $SU(2)_L\times U(1)_Y$.}
\label{tab:0}
\end{table}
\end{center}
\end{widetext}

In this section, we review our model, where we add quintet boson with zero hypercharge, that is symbolized by $\Phi_5$, to field contents of previous model ref.~\cite{Nomura:2017abu}.
In the model, several large $SU(2)_L$ multiplet fields are introduced such as septet scalar $\Phi_7$, quadruplet scalar $\Phi_4$ and quintet fermion $\Sigma_R$ 
whose hypercharges are $1$, $1/2$ and $0$,respectively.
We summarize the new field contents and their charges with the leptons and the SM Higgs field in Table~\ref{tab:0}.
In our extended model, thanks to the existence of $\Phi_5$, we find quadruplet boson $\Phi_4$ can be inert and hence the neutrino mass matrix is induced not at tree level but one-loop level. 
Note also that the quintet fermion $\Sigma_R$ is the same one as discussed in ref.~\cite{Cirelli:2005uq} which can be a dark matter candidate without imposing any additional symmetry.
However, in our case, the lightest component of $\Sigma_R$ would decay because of an interaction associated with $\Phi_4$ and leptons, and thus we do not discuss dark matter in this paper.
Here we write these multiplets by components such as,
\begin{align}
&\Phi_7 = \left( \phi^{4+}, \phi^{3+}, \phi^{++}_2, \phi^{+}_2, \phi^{0}, \phi^{-}_1, \phi^{--}_1 \right)^T, \nn \\
& \Phi_5 = \left( \xi^{++}_2, \xi^{+}_2, \xi^{0}, \xi^{-}_1 , \xi^{-}_2 \right)^T, \nn \\
& \Phi_4 = \left( \varphi^{++}, \varphi^{+}_2, \varphi^{0}, \varphi^{-}_1 \right)^T, \nn \\
& \Sigma_R = \left[ \Sigma_R^{++}, \Sigma^{+}_R, \Sigma_R^{0}, \Sigma^{+c}_L, \Sigma_L^{++c} \right]^T,
\end{align}
where subscripts for scalar components distinguish different particle with same electric charge.

First of all,  we will investigate a hierarchy among vacuum expectation values (VEVs) of the bosons.
Here, we assume that only neutral components of $H$, $\Phi_5$, and $\Phi_7$ have nonzero VEVs, which are respectively symbolized by $v/\sqrt2$ and $v_5/\sqrt{2}$ and $v_7/\sqrt{2}$.
Then, the VEVs are constrained by the $\rho$ parameter at tree-level, which is given by~\cite{Agashe:2014kda}:
\begin{align}
\rho=\frac{v^2+12 v_5^2 + 22 v_7^2}{v^2 + 4 v_7^2},
\end{align}
where the experimental value is $\rho=1.00039\pm0.00019$ at $1\sigma$ confidential level.
Also we have to satisfy the condition $v_{\rm SM}=\sqrt{v^2+12 v_5^2 + 22 v_7^2} \simeq 246$ GeV
which is the VEV of the SM-like Higgs.
Requiring these two conditions fixed by $\rho=1.00058$, we find a typical scale of solution as follows:
\begin{align}
v_7 \approx  0.748 \ {\rm GeV}, \quad v_5 \approx  1.44 \ {\rm GeV} , \quad v \approx  246 \ {\rm GeV}.
\end{align}
It suggests that small VEVs of $\Phi_5$ and $\Phi_7$ can be obtained naturally. \\
Next task is how to realize inert feature of $\Phi_4$.
The scalar potential in the model is given by
\begin{align}
\mathcal{V} = -\mu_H^2 H^\dagger H + M_4^2 \Phi_4^\dagger \Phi_4 + M_5^2 \Phi_5^\dagger \Phi_5 + M_7^2 \Phi_7^\dagger \Phi_7  + \mathcal{V}_{\rm non-trivial} +\mathcal{V}_4,
\end{align}
where $\mathcal{V}_4$ is the trivial quartic term.
The nontrivial Higgs potential terms under these symmetries are given by
\begin{align}
 \mathcal{V}_{\rm non-trivial} = &
 \mu [\Phi_4 \Phi_7^* \Phi_4] +
 \lambda_0 [H^\dag \tilde\Phi_4 H H] + \lambda_1 [H^* H^* \Phi_4 \Phi_4] + \mu_B[H^* \Phi_5^* \Phi_4]  \nonumber \\
& + \lambda_{\Phi_5^3} [\Phi_5 \Phi_5 \Phi_5] + \lambda_X [H H \Phi_7^* \Phi_5] +{\rm c.c.},
\label{Eq:lag-flavor}
\end{align}
where the first term contributes to the neutrino mass matrix at one-loop level 
and an inside bracket is ``$[ \ ]$" the $SU(2)_L$ indices, which are implicitly contracted so that it makes singlet. Then, the inert condition is given by the tadpole condition for $\Phi_4$, $\partial\mathcal{V}/\partial \Phi_4=0$;
\begin{align}
\lambda_0=-\sqrt3\frac{\mu_B v_5}{v^2}.
\end{align}
The other VEVs are also obtained by solving $\partial\mathcal{V}/\partial \Phi_5 = \partial\mathcal{V}/\partial \Phi_7 = 0$ where we omit detailed expression of them here.

\subsection{Yukawa sector}
The renormalizable Lagrangian is given by
\begin{align}
-\mathcal{L}_{Y}
&=
\sum_{\ell=e,\mu,\tau}y_{\ell} \bar L_{L_\ell} H e_{R_\ell} +(y_{\nu})_{ij} [\bar L_{L_i} \tilde\Phi_4 \Sigma_{R_j} ]
 +  (M_{R})_i [\bar \Sigma^c_{R_i} \Sigma_{R_i}] -  y_{\varphi_i} [\Phi_5 \bar \Sigma^c_{R_i} \Sigma_{R_i}] + {\rm h.c.}, 
\label{Eq:lag-flavor}
\end{align}
where $(i,j)=1-3$, and $y_\ell$ contributes to the charged-lepton masses in the SM and $M_R$, $y_\ell$, and $y_{\varphi}$ are assumed to be diagonal basis.

\subsection{Exotic fermion masses}
Here we consider masses of the extra particles in the model.
The masses for components in $\Sigma_R$ are obtained from the last two terms of Eq.~\eqref{Eq:lag-flavor} after $\Phi_5$ developing a VEV.
Then we obtain the mass terms
\begin{equation}
L_{M_\Sigma} = \left( M_R +  \frac{2v_5}{\sqrt{6} }y_\varphi \right) \bar \Sigma^{++} \Sigma^{++} + \left( M_R -  \frac{v_5}{\sqrt{6} }y_\varphi \right) \bar \Sigma^{+} \Sigma^{+} +
\left( M_R -  \frac{v_5}{\sqrt{6} }y_\varphi \right) \bar \Sigma^{0c}_R \Sigma^{0}_R,  
\end{equation}
where the mass of neutral component is Majorana type.
Since quintuplet VEV $v_5$ cannot be large, the mass differences among the components are at most several GeV.

The masses of the components of scalar multiplet $\Phi_{4,5,7}$ can obtain separate values from the contribution of non-trivial terms in the potential.
Here we consider the mass differences are at most $\mathcal{O}(100)$ GeV scale and masses for $\Phi_{4,5,7}$ are dominantly given by $M_{4,5,7}$.

\subsection{ Neutrino mass matrix}
Let us first decompose the relevant Lagrangian in order to derive the neutrino mass matrix.
The neutrino mass matrix is given in terms of $y_\nu$, and its explicit form is found as
\begin{align}
-{\cal L}&
\supset 
 \frac{ {(y_\nu)_{ij}}}{\sqrt2}
\bar\nu_{L_i} \Sigma_{R_j}^0 (\varphi_R -i \varphi_I)
+ M_{R_i} \bar\Sigma^{0c}_{R_i}\Sigma^0_{R_i} +{\rm h.c.},
\label{eq:Yukawa}
\end{align}
where the mass difference between the real  $\Phi_4$ component; $\varphi_R$ and the imaginary one; $\varphi_I$ is generated by the term $\mu$ through VEV of $\Phi_7$. 
\begin{figure}[tb]\begin{center}
\includegraphics[width=70mm]{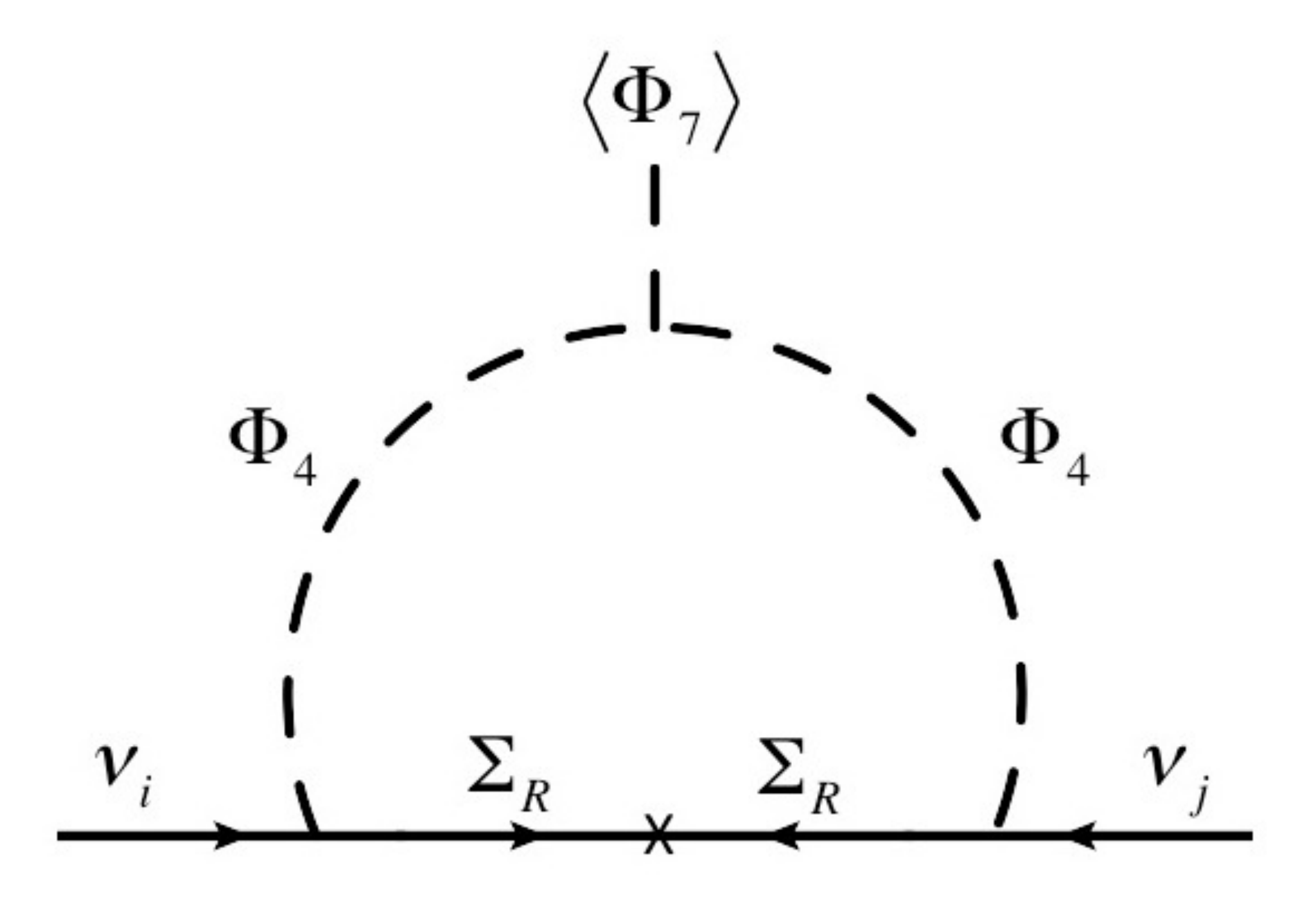}
\caption{{ Contribution to neutrino masses at one-loop level.}} \label{fig:neut1}\end{center}\end{figure}
Then the formula of active neutrino mass matrix $m_\nu$ as shown in Figure~\ref{fig:neut1} is given by 
\begin{align}
& (m_{\nu})_{ij}
=\frac{{\mu v_7}}{(4\pi)^2} \sum_{a=1}^3\frac{(y_\nu)_{ia}(y_\nu^T)_{aj}}{M_{R_a}} F_I(r_{R_a},r_{I_a}),
\nonumber \\
& F_I(r_1,r_2) =\frac{r_1 \ln[r_1]-r_2\ln[r_2]+r_1r_2\ln[r_2/r_1]}{(1-r_1)(1-r_2)},
\end{align}
where we define $r_{R(I)_a}\equiv (m_{\varphi_{R(I)}}/M_{R_a})^2$ with $a=1-3$.
The mass matrix $({m}_\nu)_{ij}$ can be generally diagonalized by the Pontecorvo-Maki-Nakagawa-Sakata mixing matrix $V_{\rm MNS}$ (PMNS)~\cite{Maki:1962mu} as
\begin{align}
({m}_\nu)_{ab} &=(V_{\rm MNS} D_\nu V_{\rm MNS}^T)_{ab},\quad D_\nu\equiv (m_{\nu_1},m_{\nu_2},m_{\nu_3}),
\\
V_{\rm MNS}&=
\left[\begin{array}{ccc} {c_{13}}c_{12} &c_{13}s_{12} & s_{13} e^{-i\delta}\\
 -c_{23}s_{12}-s_{23}s_{13}c_{12}e^{i\delta} & c_{23}c_{12}-s_{23}s_{13}s_{12}e^{i\delta} & s_{23}c_{13}\\
  s_{23}s_{12}-c_{23}s_{13}c_{12}e^{i\delta} & -s_{23}c_{12}-c_{23}s_{13}s_{12}e^{i\delta} & c_{23}c_{13}\\
  \end{array}
\right],
\end{align}
where we neglect the Majorana phase as well as Dirac phase $\delta$ in the numerical analysis for simplicity.
The following neutrino oscillation data at 95\% confidential level~\cite{Agashe:2014kda} is given as
\begin{eqnarray}
&& 0.2911 \leq s_{12}^2 \leq 0.3161, \; 
 0.5262 \leq s_{23}^2 \leq 0.5485, \;
 0.0223 \leq s_{13}^2 \leq 0.0246,  
  \\
&& 
  \ |m_{\nu_3}^2- m_{\nu_2}^2| =(2.44\pm0.06) \times10^{-3} \ {\rm eV}^2,  \; 
  \ m_{\nu_2}^2- m_{\nu_1}^2 =(7.53\pm0.18) \times10^{-5} \ {\rm eV}^2. \nn
  \label{eq:neut-exp}
  \end{eqnarray}
%
The observed PMNS matrix can be realized by introducing the following parametrization.
Here we parametrize the Yukawa coupling by $y_\nu${, so called Casas-Ibarra parametrization~\cite{Casas:2001sr},} as follows
\begin{align}
y_\nu
&= V_{\rm MNS} \sqrt{D_\nu} OR^{-1/2},
\label{yl-sol}
\\
R_{aa}&\equiv
\frac{{\mu v_7}}{(4\pi)^2} \sum_{a=1}^3\frac{1}{M_{R_a}}  F_I(r_{R_a},r_{I_a}),
 \label{R-sol}
\end{align}
where $O$ is an arbitrary complex orthogonal matrix with three degrees of freedom.

\subsection{LFVs and muon $g-2$}
LFVs at one-loop level arise from $y_\nu$, and the related Lagrangian is given by
\begin{align}
-{\cal L}&
 \supset (y_\nu)_{ij}\bar\ell_{L_i} \left(\frac{1}{\sqrt2}\Sigma_{R_j}^0 \varphi_1^-  + \frac12 \Sigma_{1R_j}^+ \varphi^{--} +  \frac{\sqrt3}{2}\Sigma_{2R_j}^- \varphi^{0*} +  \Sigma_{2R_j}^{--} \varphi_{2}^{+}  \right) +{\rm h.c.}.
\label{eq:Yukawa}
\end{align}
Then the branching ratio is found as
\begin{align}
&{\rm BR}(\ell_i\to \ell_j)\approx
\frac{75}{64\pi}\frac{\alpha_{em} C_{ij}}{G_F^2}|y_{\nu_{ja}} y^\dag_{\nu_{ai}} G(\varphi,\Sigma_a)|^2\left(1+\frac{m^2_{\ell_j}}{m^2_{\ell_i}}\right)^2,\\
&G(m_1,m_2)=\int_0^1dx\int_0^{1-x-y}dy\frac{xy}{(x^2-x)m^2_{\ell_i}+x m_1^2+(1-x) m_2^2},
\end{align}
where $C_{21}=1$, $C_{31}=0.1784$, $C_{32}=0.1736$, $\alpha_{em}(m_Z)=1/128.9$, and $G_F=1.166\times10^{-5}$ GeV$^{-2}$.
The experimental upper bounds are given by~\cite{TheMEG:2016wtm, Aubert:2009ag,Renga:2018fpd}
\begin{align}
{\rm BR}(\mu\to e\gamma)\lesssim 4.2\times10^{-13},\quad 
{\rm BR}(\tau\to e\gamma)\lesssim 3.3\times10^{-8},\quad
{\rm BR}(\tau\to\mu\gamma)\lesssim 4.4\times10^{-8},
\label{eq:lfvs-cond}
\end{align}
which will be imposed in our numerical calculation.

Muon $g-2$ is also induced from the same term and is given by
\begin{align}
\Delta a_\mu= \frac{5m^2_\mu}{(4\pi)^2} \sum_a y_{\nu_{2a}} y^\dag_{\nu_{a2}} G(\varphi,\Sigma_a) ,
\end{align}
while the experimental result implies $\Delta a_\mu=(26.1\pm 8.0)\times 10^{-10}$~\cite{Hagiwara:2011af}.

{
\subsection{Beta functions of $g$ and $g_Y$}
\label{beta-func}
Here we discuss running of gauge couplings and estimate the effective energy scale by evaluating the Landau poles for $g$ and $g_Y$ in the presence of new fields 
with nonzero multiple hypercharges.
Each of the new beta function of $g$ and  $g_Y$ for one $SU(2)_L$ quintet fermion ($\Sigma_R$), quartet boson ($\Phi_4$), {quintet boson $\Phi_5$,}
and septet boson($\Phi_7$) with $(0,1/2,1)$ hypercharge is given by
\begin{align}
\Delta b^{\Sigma_R}_g=\frac{20}{3}, \quad \Delta b^{\Phi_4}_g=\frac{5}{3} \ , \quad \Delta b^{\Phi_5}_g=\frac{10}{3} 
\ ,\quad\Delta b^{\Phi_7}_g=\frac{28}{3} ,\\
\Delta b^{\Sigma_R}_Y=0, \quad \Delta b^{\Phi_4}_Y=\frac{3}{5} \, \quad \Delta b^{\Phi_5}_Y=0\ ,\quad\Delta b^{\Phi_7}_Y=\frac{7}{5}.
\end{align}
Then one finds the energy evolution of the gauge coupling $g$ and $g_Y$ as~\cite{Kanemura:2015bli}
\begin{align}
\frac{1}{g^2(\mu)}&=\frac1{g^2(m_{in.})}-\frac{b^{SM}_g}{(4\pi)^2}\ln\left[\frac{\mu^2}{m_{in.}^2}\right]\nn\\
&
-\theta(\mu-m_{th.}) \frac{\Delta b^{\Sigma_R}_g}{(4\pi)^2}\ln\left[\frac{\mu^2}{m_{th.}^2}\right]
-\theta(\mu-m_{th.}) \frac{\Delta b^{\Phi_4}_g+\Delta b^{\Phi_5}_g+\Delta b^{\Phi_7}_g}{(4\pi)^2}\ln\left[\frac{\mu^2}{m_{th.}^2}\right],\label{eq:rge_g}\\
\frac{1}{g^2_Y(\mu)}&=\frac1{g_Y^2(m_{in.})}-\frac{b^{SM}_Y}{(4\pi)^2}\ln\left[\frac{\mu^2}{m_{in.}^2}\right]
-\theta(\mu-m_{th.}) \frac{\Delta b^{\Phi_4}_Y+\Delta b^{\Phi_7}_Y}{(4\pi)^2}\ln\left[\frac{\mu^2}{m_{th.}^2}\right],\label{eq:rge_gy}
\end{align}
where $\mu$ is a reference energy, $b^{SM}_Y=41/6$, $b^{SM}_g=-19/6$, and we assume to be $m_{in.}(=m_Z)<m_{th.}=$500 GeV, being respectively threshold masses of exotic fermions and bosons for $m_{th.}$.
The resulting flow of $g_Y(\mu)$ is then given by the Figure~\ref{fig:rge} for $g$ for each of $m_{in.}=(0.5,1,10)$ TeV,
where $g_Y$ is valid up to Planck scale.
This figure shows that $g$ is relevant up to the mass scale $\mu\approx (2,4)\times10^2$ TeV for  $m_{in.}=(0.5,1)$ TeV and  $\mu\approx 5\times10^3$ TeV for  $m_{in.}=10$ TeV.
Thus our theory does not spoil, as far as we work on at around the scale of TeV.

\begin{figure}[tb]
\begin{center}
\includegraphics[width=13cm]{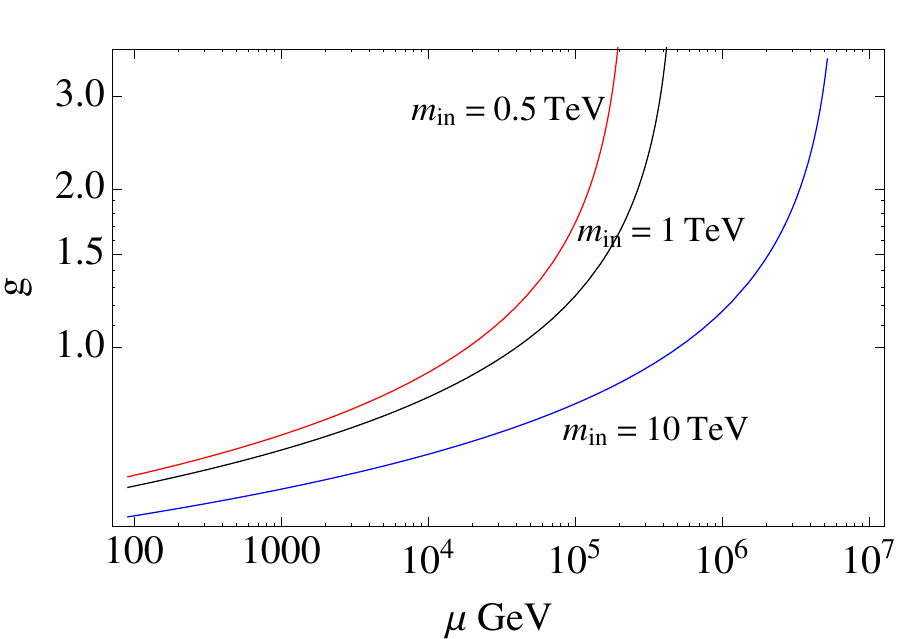}
\caption{The running of $g$  in terms of  $\mu$, depending on $m_{in.}=(0.5,1)$ TeV}
\label{fig:rge}
\end{center}\end{figure}
}

\section{Numerical analysis and Implications to physics at the LHC}

In this section, we perform numerical analysis of neutrino mass matrix taking into account LFV constraints.
Then implications to collider physics are discussed adopting benchmark point accommodating with neutrino data and the constraints.

\subsection{Numerical analysis for neutrino sector}

Here we carry out numerical analysis scanning free parameters and check if we can fit the neutrino data.
The free parameters are chosen within the range of 
\begin{align}
& M_{R_1} \in [500, 2000] \ {\rm GeV}, \ M_{R_{2,3}} \in [M_{R_1}, 5000] {\rm GeV}, \nonumber \\
& m_{\varphi_R} \in [100, 2000] \ {\rm GeV}, \ m_{\varphi_I} \in [m_{\varphi_R} - 1, m_{\varphi_R} + 1] \ {\rm GeV}, \ \mu \in [0.3, 0.4] \ {\rm GeV},
\end{align}
where we calculate $y_\nu$ as output using Eq.~\eqref{yl-sol} scanning parameters in orthogonal matrix $O$ as $\mathcal{O}(0.1)$--$\mathcal{O}(1)$ randomly. 
We find that the neutrino data can be fitted with $\mathcal{O}(0.1)$ to $\mathcal{O}(1)$ Yukawa couplings $(y_\nu)_{ij}$ satisfying constrains from LFV processes.
In addition, muon $g-2$ is found to be maximally $\sim 4 \times 10^{-11}$ due to constraint from $\mu \to e \gamma$.
We will choose some benchmark points satisfying all the constraints and provide maximal muon $g-2$ to consider collider physics in the following subsection.

\subsection{Implications to LHC physics}

In this subsection, we discuss collider physics regarding the charged particles from large fermion/scalar multiplets in the model.
In particular, we focus on particles from quintuplet fermion $\Sigma_R$ and $\Phi_4$ which propagate inside the loop diagram in neutrino mass generation 
where we assume components from $\Phi_7$ and $\Phi_5$ are heavier than those particles.
The relevant gauge interactions associated with $\Sigma_R$ can be written by
\begin{align}
\bar \Sigma_R \gamma^\mu i D_\mu \Sigma_R \supset  &  \bar \Sigma^{++} \gamma^\mu \left( 2 e A_\mu + 2 g c_W  Z_\mu  \right) \Sigma^{++} + 
\bar \Sigma^{+} \gamma^\mu \left( e A_\mu + g c_W  Z_\mu  \right) \Sigma^{+} \nonumber \\
&  - \sqrt{2} g   \bar \Sigma^{++} \gamma^\mu W_\mu^+  \Sigma^{+} - \sqrt{3} g   \bar \Sigma^{+} \gamma^\mu W_\mu^+ \Sigma_R^{0} - \frac{\sqrt{5} g}{\sqrt{2}}   \bar \Sigma^{+} \gamma^\mu W_\mu^+ \Sigma^{0c}_R \nonumber \\
& - \sqrt{2} g   \bar \Sigma^{+} \gamma^\mu W_\mu^-  \Sigma^{++} - \sqrt{3} g   \bar \Sigma^{0}_R \gamma^\mu W_\mu^- \Sigma^{+} - \frac{\sqrt{5} g}{\sqrt{2}}   \bar \Sigma^{0 c}_R \gamma^\mu W_\mu^- \Sigma^{+},
\label{eq:ke1}
\end{align}
where $s_W(c_W) = \sin \theta_W (\cos \theta_W)$ with the Weinberg angle $\theta_W$.
Also the relevant gauge interactions associated with $\Phi_4$ can be obtained from following kinetic term
\begin{align}
|D_\mu \Phi_4|^2 =  &
\sum_{m= -\frac32,-\frac12,\frac12,\frac32} \biggl|  \left[ \partial_\mu - i \left(\frac12+m \right) e A_\mu - i \frac{g}{c_W} \left(m - \left( \frac12+m \right) s_W^2 \right) Z_\mu \right] (\Phi_4)_{m} \nonumber \\
& \qquad \qquad \qquad + \frac{ig}{\sqrt{2}} \sqrt{ \left(\frac32 + m \right) \left(\frac52 -m \right) } W^+_\mu (\Phi_4)_{m-1} \nonumber \\
& \qquad \qquad  \qquad + \frac{ig}{\sqrt{2}} \sqrt{ \left(\frac32 - m \right) \left(\frac52 +m \right) } W^-_\mu (\Phi_4)_{m+1} \biggr|^2 \\
& \supset i \sqrt{\frac32} g W^-_\mu (\partial^\mu \varphi^-_2 \varphi^{++} - \partial^\mu \varphi^{++} \varphi^-_2) 
+ i \sqrt{2} g W^-_\mu ( \partial^\mu \varphi^{0*} \varphi^+_2 - \partial^\mu \varphi_2^+ \varphi^{0*} ) + h.c.,
\label{eq:ke2}
\end{align}
where $(\Phi_4)_m$ indicates the component of $\Phi_4$ which has the eigenvalue of diagonal $SU(2)$ generator $T_3$ given by $m$, 
and the last line shows the relevant interactions for decay of $\varphi^{\pm \pm}$ and $\varphi^\pm_2$.
In addition, Yukawa coupling associated with $\Sigma_R$ can be expanded as 
\begin{align}
-{\cal L}& \supset (y_\nu)_{ij}
\left[
\bar\nu_{L_i} \left(\frac{1}{\sqrt2}\Sigma_{R_j}^0 \varphi^{0*} + \frac{\sqrt3}{2} \Sigma_{R_j}^+ \varphi_1^- -  \frac{1}{2} \Sigma_{L_j}^{+c} \varphi_{2}^+ +  \Sigma_{R_j}^{++} \varphi^{--}  \right)\right.
\nn\\
&\left.  \qquad \qquad +
\bar\ell_{L_i} \left(\frac{1}{\sqrt2}\Sigma_{R_j}^0 \varphi_1^-  + \frac12 \Sigma_{R_j}^+ \varphi^{--} - \frac{\sqrt3}{2}\Sigma_{L_j}^{+c} \varphi^{0*} +  \Sigma_{L_j}^{++c} \varphi_{2}^{+}  \right)\right] \nn \\
& + \frac{(y_\varphi)_{ij}}{\sqrt{6}} \left[  \bar \Sigma^{0c}_{R_i} \Sigma^{0}_{R_j} \xi_0 + \bar \Sigma^{+c}_{R_i} \Sigma^{+c}_{L_j} \xi_0 - 2 \bar \Sigma^{++c}_{R_i} \Sigma^{++c}_{L_j} \xi_0 
- \sqrt{\frac32} \bar \Sigma^{+}_{L_i} \Sigma^{+c}_{L_j} \xi_1^{++} + \sqrt{6} \bar \Sigma^{++c}_{R_i} \Sigma^{+c}_{L_j} \xi_2^- \right. \nn \\
& \left. \qquad \qquad - \sqrt{\frac32} \bar \Sigma^{+c}_{R_i} \Sigma^{+}_{R_j} \xi_2^{--} - \bar \Sigma^{0c}_{R_i} \Sigma^{+c}_{L_j} \xi_1^+ +2 \bar \Sigma^{0c}_{R_i} \Sigma^{++c}_{L_j} \xi_2^{++} + 
\bar \Sigma^{0c}_{R_i} \Sigma^{+}_{R_j} \xi_2^- + 2  \bar \Sigma^{0c}_{R_i} \Sigma^{++}_{R_j} \xi_2^{--}  \right] \nn \\
& +{\rm h.c.},
\label{eq:Yukawa}
\end{align}
where these terms are obtained  from second and forth term in Eq.~\eqref{Eq:lag-flavor}.

Here we consider signature of the model focusing on the production of doubly charged fermion in $\Sigma_R$.
The doubly charged fermion pair can be produced via electroweak interaction as 
\begin{equation}
p p \to Z/\gamma \to \Sigma^{++} \Sigma^{--}.
\end{equation}
\begin{figure}[tb]
\begin{center}
\includegraphics[width=8.0cm,height=7.5cm]{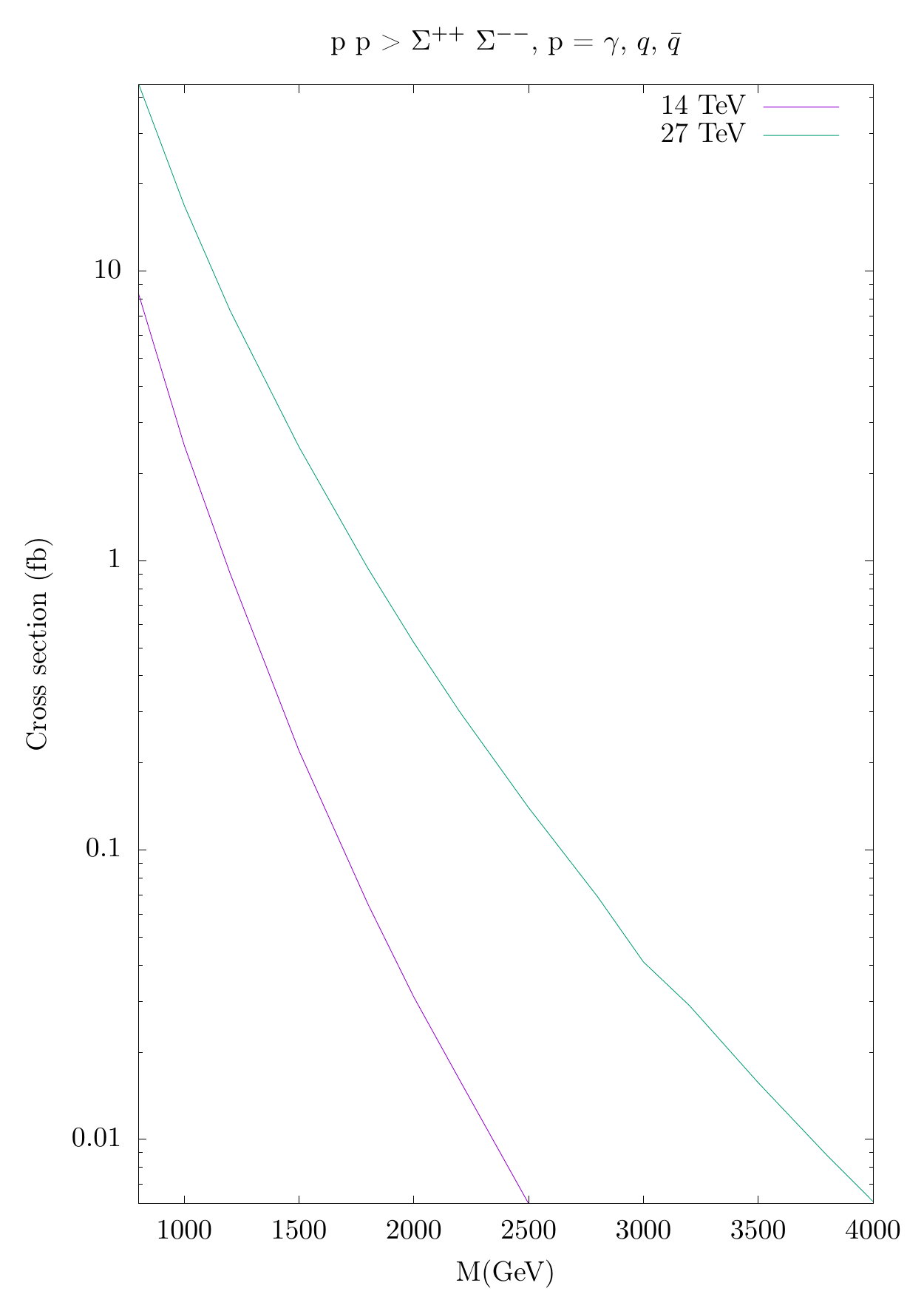}
\caption{The cross section for pair production process $p p \to Z/\gamma \to \Sigma^{++} \Sigma^{--}$ as a function of $\Sigma^{\pm \pm}$ mass at 14 TeV and 27 TeV.}
\label{fig:cross1}
\end{center}
\end{figure}
In order to compute the cross-sections and generate events at the LHC, 
we incorporate the model Lagrangian of Eq.~(\ref{eq:ke1}), Eq.~(\ref{eq:ke2}), 
and Eq.~(\ref{eq:Yukawa}) in FeynRules (v2.3.13) 
\cite{Alloul:2013bka,Christensen:2008py}. Using FeynRules, we generate the 
model file for MadGraph5\_aMC@NLO (v2.2.1)~\cite{Alwall:2014hca}. For the 
cross-sections, we use {the} NNPDF23LO1 parton distributions~\cite{Ball:2012cx} 
with the factorization and renormalization scales at the 
central $m_T^2$ scale after $k_T$-clustering of the event.
We have computed the signal cross section of $p p \to Z/\gamma \to \Sigma^{++} \Sigma^{--}$, 
where $p = q, \bar q, \gamma$. The cross sections are normalised to the 5 flavor scheme.
We have shown the production cross section in Figure~\ref{fig:cross1}.
The inclusion of the photon PDF increases the signal cross section significantly as the coupling 
is proportional to the charge of the fermion. Moreover, 
inclusion of photon PDF is important for the consistency of the calculation 
as the other PDF's are determined up to NNLO in QCD. We would like to note that, in 
view of the above, NNPDF \cite{Ball:2014uwa,Ball:2013hta}, MRST \cite{Martin:2004dh} 
and CTEQ \cite{Schmidt:2015zda} have already included photon PDF into their PDF sets.

The possible decay modes of the fermions are,
\begin{align}
& \Sigma^{\pm \pm} \to \ell^\pm \varphi^{\pm }_2 \to \ell^\pm W^{\pm*} \varphi_R \to \ell^\pm W^{\pm*} h h, \nn \\
& \Sigma^{\pm \pm} \to \nu \varphi^{\pm \pm} \to \nu W^{\pm*} \varphi^+_2 \to \nu W^{\pm*}  W^{\pm*}  \varphi_R \to \nu W^{\pm*}  W^{\pm*}  h h,
\end{align}
where $\varphi_R \to hh$ decay is induced by the interaction with coupling $\lambda_0$.
This gives rise to final states comprising of a number of leptons, jets, and missing energy.

Some benchmark points of the model are shown in (Table \ref{tab:1}) for 
various values of the parameter space. Note that the values of $(y_\nu)$ satisfies 
the constrains from LFV's and provide maximal muon $g-2$ contribution, as discussed in the previous 
subsection. It is evident from Table \ref{tab:1} that the branching ratio 
of $\Sigma^{\pm \pm}$ to $e$, $\mu$ and $\tau$ depends on the choice of 
$y_\nu$. In Table \ref{tab:1}, branching ratio to ($l=e, \mu$) and $\tau$ are 
given separately. In the collider analysis 
we focus on states involving $l=e, \mu$ only and we have assumed a simplified scenario, where 
BR($\Sigma^{\pm \pm}\to \ell^\pm \varphi^{\pm }_2)
=(\Sigma^{\pm \pm} \to \nu \varphi^{\pm \pm})\sim 50\%$ and we assumed it to be same for every 
lepton family. A detailed analysis involving the parameter space where the decay of 
$\Sigma^{\pm \pm}$ to ($\tau \varphi^{\pm \pm}$) is maximum, will be studied elsewhere.
\begin{table}
\centering
\begin{tabular}{|c|c|c|c|}
\hline\hline
-&BP1 & BP2& BP3 \\
\hline\hline
$v_5$ & 1.44&1.44 &1.44\\
$v_7$&0.748 &0.748 &0.748\\
$\lambda_0$ (GeV)& 0.01 & 0.01& 0.01\\
$(y\nu)_{11}$&  $-0.424903 - 0.433832 i$& $-0.211747 + 0.0786788 i$& $0.451545 + 0.281382 i $\\
$(y\nu)_{21}$& $-0.515018 + 0.294156 i$& $-0.174239-0.404063 i$& $-0.455368 - 0.660787 i$ \\
$(y\nu)_{31}$&  $-0.674845 + 0.282308 i$ &  $0.218317 - 0.045373 i$& $1.17154 - 0.229787 i$ \\
BR$(\Sigma^{\pm \pm} \to (l^\pm \varphi^{\pm \pm})$&0.28 &0.43 &0.20\\
BR$(\Sigma^{\pm \pm} \to (\tau^\pm \varphi^{\pm \pm})$&0.22&0.07&0.30\\
\hline\hline
\end{tabular}
\caption{Different benchmark points of this model. $\lambda_0 = -4.12\times 10^{-5}\mu_{B}$, 
where we have kept $\mu_B$ at EW scale, thus keeping $\lambda_0$ well within perturbative limit.}
\label{tab:1}
\end{table}

Once $\Sigma^{\pm \pm}$ is produced in pair, the three major channels 
to observe this signal are ($l^+W^+ h h$,$l^-W^- h h$), ($W^+W^+ h h$,$W^-W^- h h$) + MET 
and ($l^+W^+ h h$,$W^-W^- h h$) + MET. $W$ can decay either leptonically 
with BR $(W^\pm \to l \nu)$ = 0.108 for each lepton or
hadronically with BR $(W^\pm \to q \bar q)$ = 0.676. The cross section$\times$ BR 
in each possible case is given below for two cases, where all $W$'s decay 
leptonically or all decay hadronically because that will give the minimum 
$\sigma \times BR$ and maximum $\sigma\times BR$ respectively. There can be many 
other possible channels with different combinations 
of leptons and jets with $\sigma\times BR$ varying between these two numbers. 
We have kept the mass of $\Sigma^{\pm \pm}$ at 1 TeV in the following.

$\sigma\times BR$$(l^+W^+ h h$, $l^-W^- h h)$ $\to$ $(l^+l^+)(l^-l^-)(hhhh)+MET$ $\sim$ 0.06 fb\\
$~~~~~~~~~~~~~~~~~~~~~~~~~~~~~~~~~~~~~~~~~~~\to$ $(l^+l^+)(4j)(hhhh)$ $~~~~~~~~~~~~~\sim$ 0.58 fb\\

$\sigma\times BR$$(W^+W^+ h h$, $W^-W^- h h$+ MET) $\to$ $(l^+l^+)(l^-l^-)(hhhh)+MET$ $\sim$ 0.003 fb\\
$~~~~~~~~~~~~~~~~~~~~~~~~~~~~~~~~~~~~~~~~~~~~~~~~~~~~~~~\to$ $(l^+l^+)(8j)(hhhh)+ MET$ $~~~~\sim$ 0.26 fb\\

$\sigma\times BR$$(l^+W^+ h h$, $W^-W^- h h$ + MET) $\to$ $(l^+l^+)(l^-l^-)(l^-l^-)(hhhh)+MET$ $\sim$ 0.03 fb\\
$~~~~~~~~~~~~~~~~~~~~~~~~~~~~~~~~~~~~~~~~~~~~~~~~~~~~~~~~\to$ $(l^+)(6j)(hhhh)+ MET$ $~~~~~~~~~~~~~\sim$ 0.77 fb\\

If the final state is rich with jets, coming from the decay 
of $W$, $\sigma\times BR$ is higher, but 
then the QCD background will be dominant. On the other hand, final states with the requirement 
of 1-4 leptons have negligible background and the channels are comparatively clean. 
Hence here we focus on the final state $(l^+l^+)(l^-l^-)(hhhh)+MET$ as the cross section is relatively 
high, compared to the other leptonic channels. 

In the final state we require at least 4 b-jets, coming from the Higgs and 
at least one of the two oppositely charged lepton pairs. We have checked that if we demand $\geq$6 b-jets, 
or exactly two oppositely charged lepton pairs, the signal 
efficiency decreases significantly. For event reconstruction 
we have based our analysis on ref.~\cite{Aaboud:2018knk}. Events with 
b-tagged jet with transverse momentum $p_T>$ 40 GeV and $|\eta|<2.5$ are considered.
Then we select at least two Higgs boson candidates, each
composed of two b-tagged anti-$k_t$ small-$R$ jets, with invariant masses near $m_H$.
The invariant mass of the two-Higgs-boson-candidate system $M_{4b}$ 
is used as the final discriminant which is expected to peak at $m_{{\color{red} \varphi}_R}$.
For the leading and subleading leptons, the transverse momentum selections 
are $p_T(l_1)>$ 40 GeV, $p_T(l_2)>$20 GeV, if any 
other lepton is present in the event then the minimum $p_T$ is required to be 
10 GeV. The other selections for the leptons are $|\eta|<2.5$, 
$\Delta R(l,l) > 0.4$, $\Delta R(l,b) > 0.4$.

\begin{figure}[tb]
\begin{center}
\includegraphics[width=6.9cm,height=5.7cm]{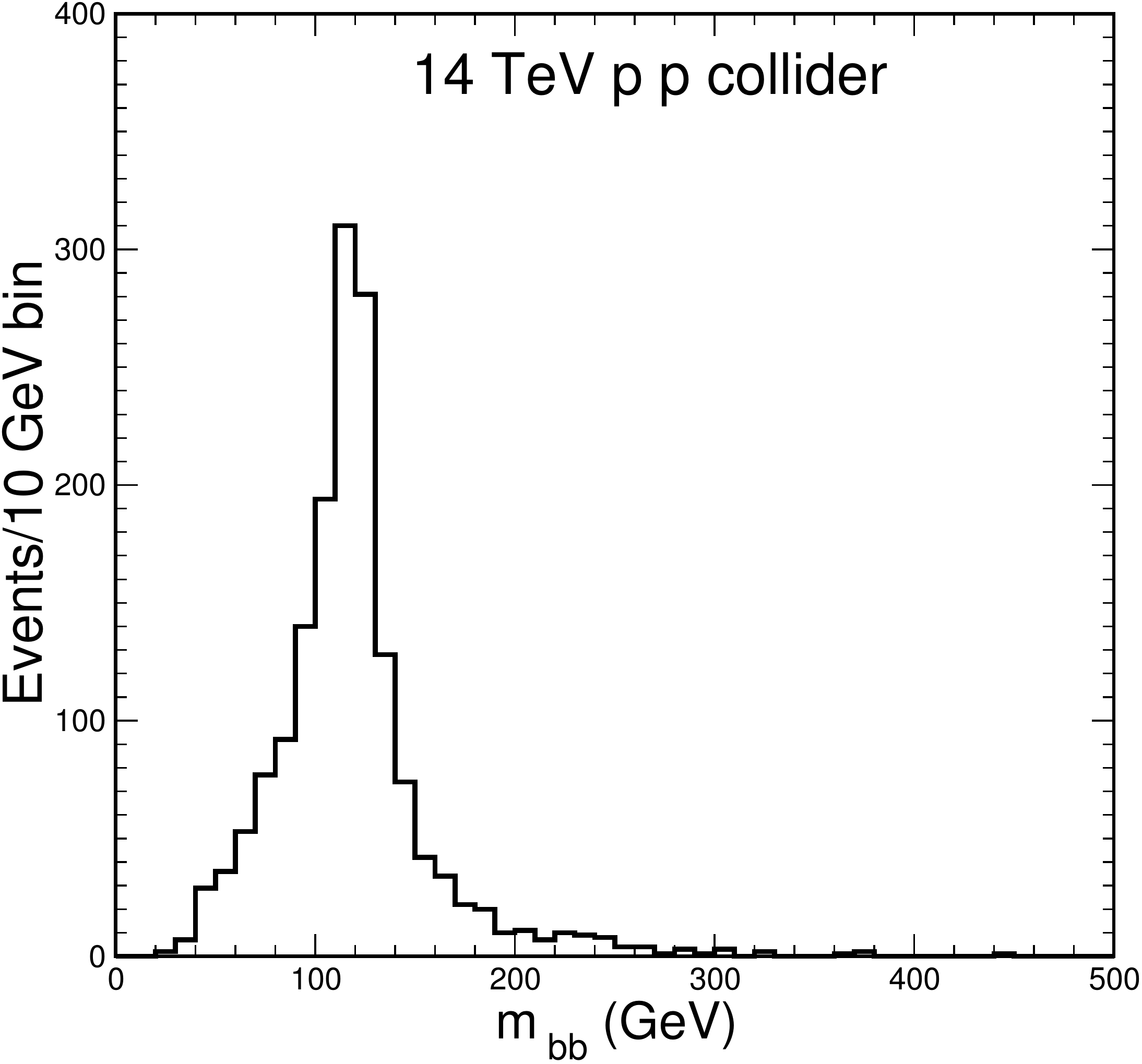}
\includegraphics[width=6.9cm,height=5.7cm]{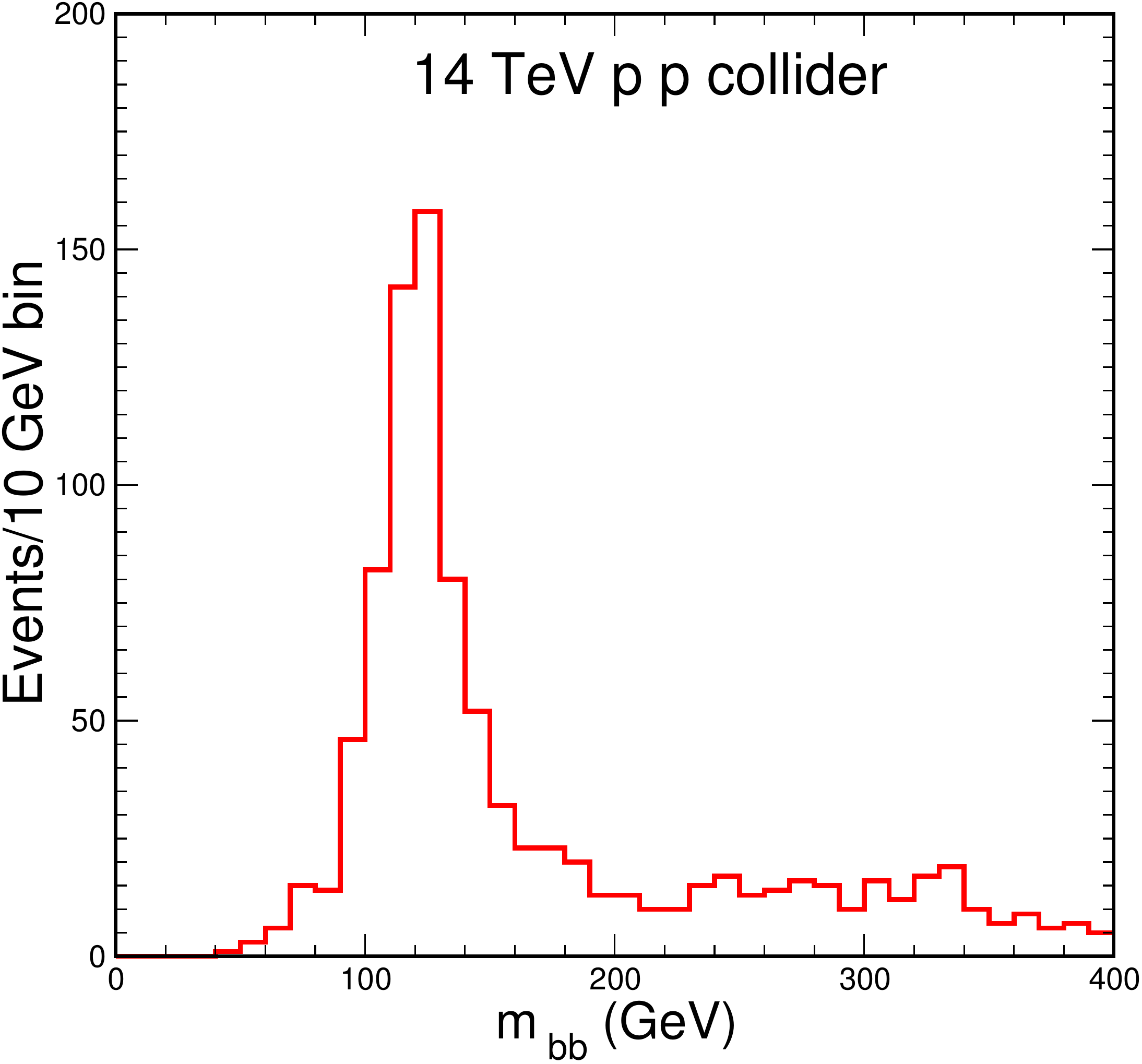}
\caption{The invariant mass distribution of leading (left) and subleading (right) 
b-jet pairs in events of $p p \to Z/\gamma \to \Sigma^{++} \Sigma^{--}$ at 14 TeV.}
\label{fig:2}
\end{center}
\end{figure}

\begin{figure}[tb]
\begin{center}
\includegraphics[width=7.9cm,height=6.7cm]{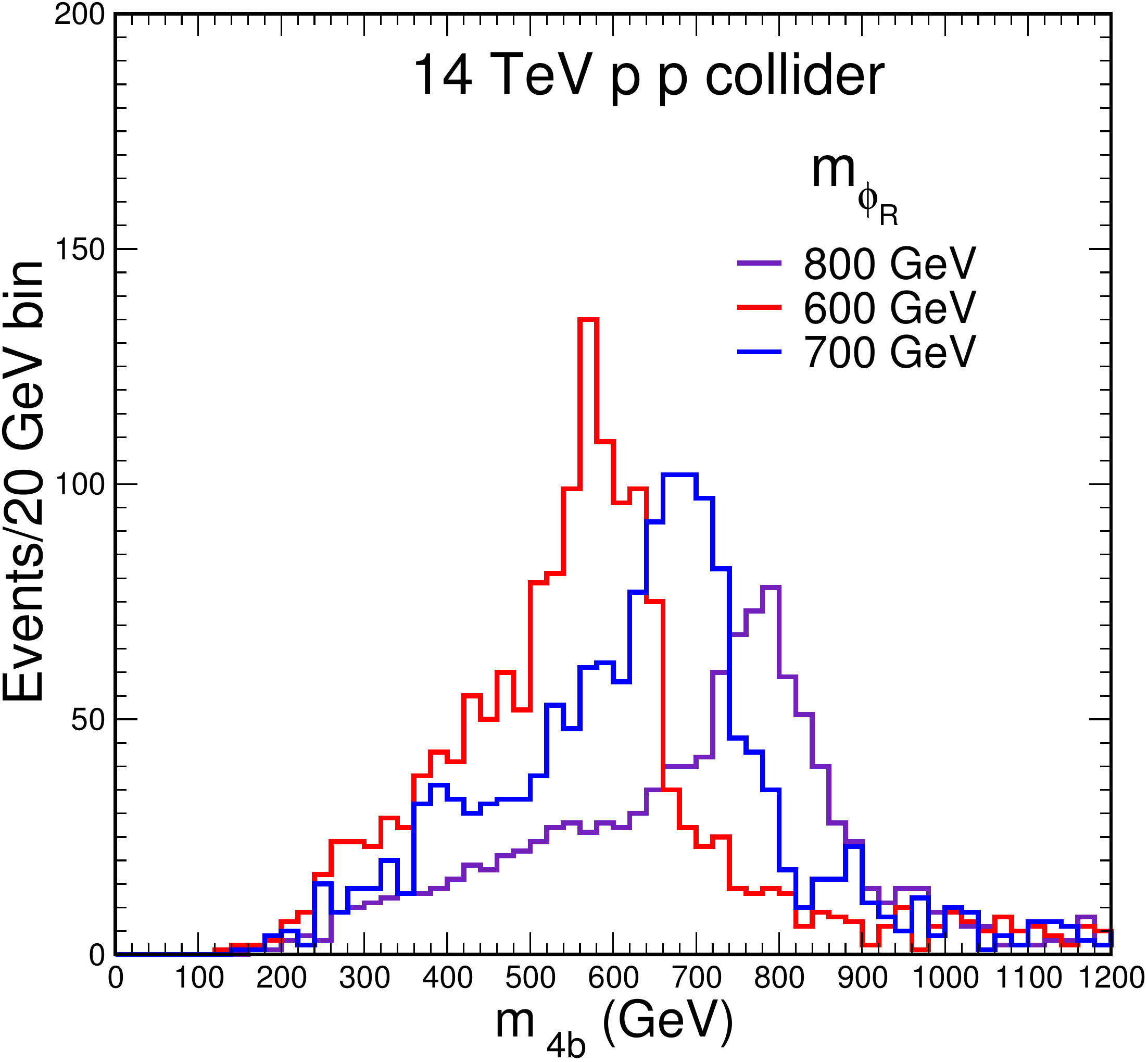}
\caption{The four b-jet invariant mass distribution in events of $p p \to Z/\gamma \to \Sigma^{++} \Sigma^{--}$ at 14 TeV for different masses of $\varphi_R$.}
\label{fig:3}
\end{center}
\end{figure}

In our study we choose three representative points, $m_{\varphi_R}$ = 600, 700 and 800 GeV,
where the mass difference between the other scalars and the fermions of the model are 
kept well with O(10) GeV to O(100) GeV. Pairing of the b-jets is required to satisfy 
the following criteria for the angular distance between them, (see ref.~\cite{Aaboud:2018knk}).\\
\begin{center}
$\frac{360}{m_{4b}}-0.5 < \Delta R_{bb}({\rm leading}) < \frac{653}{m_{4b}}+0.475 $\\
$\frac{235}{m_{4b}} < \Delta R_{bb}({\rm subleading}) < \frac{875}{m_{4b}}+0.35 $
\end{center}
From the first combination we get the leading Higgs and from the 
second we get the subleading Higgs candidate. Moreover, the pair 
that gives $m_{2b}$ closest to the Higgs mass are considered to 
be the leading pair of jets. In order to reject multijet events 
we also choose $|\Delta\eta(hh)|<1.5$.
The distribution of the leading and sub-leading b-jet pairs are 
shown in Figure~\ref{fig:2}.
We also show the 4-bjet invariant mass distribution 
in Figure~\ref{fig:3} for $m_{\varphi_R}$ = 600, 700 and 800 GeV.
If the mass of $m_{\varphi_R}\geq$ 1 TeV, then boosted jet techniques 
are required for the analysis which is beyond the scope of this paper.

\begin{figure}[tb]
\begin{center}
\includegraphics[width=7.9cm,height=6.7cm]{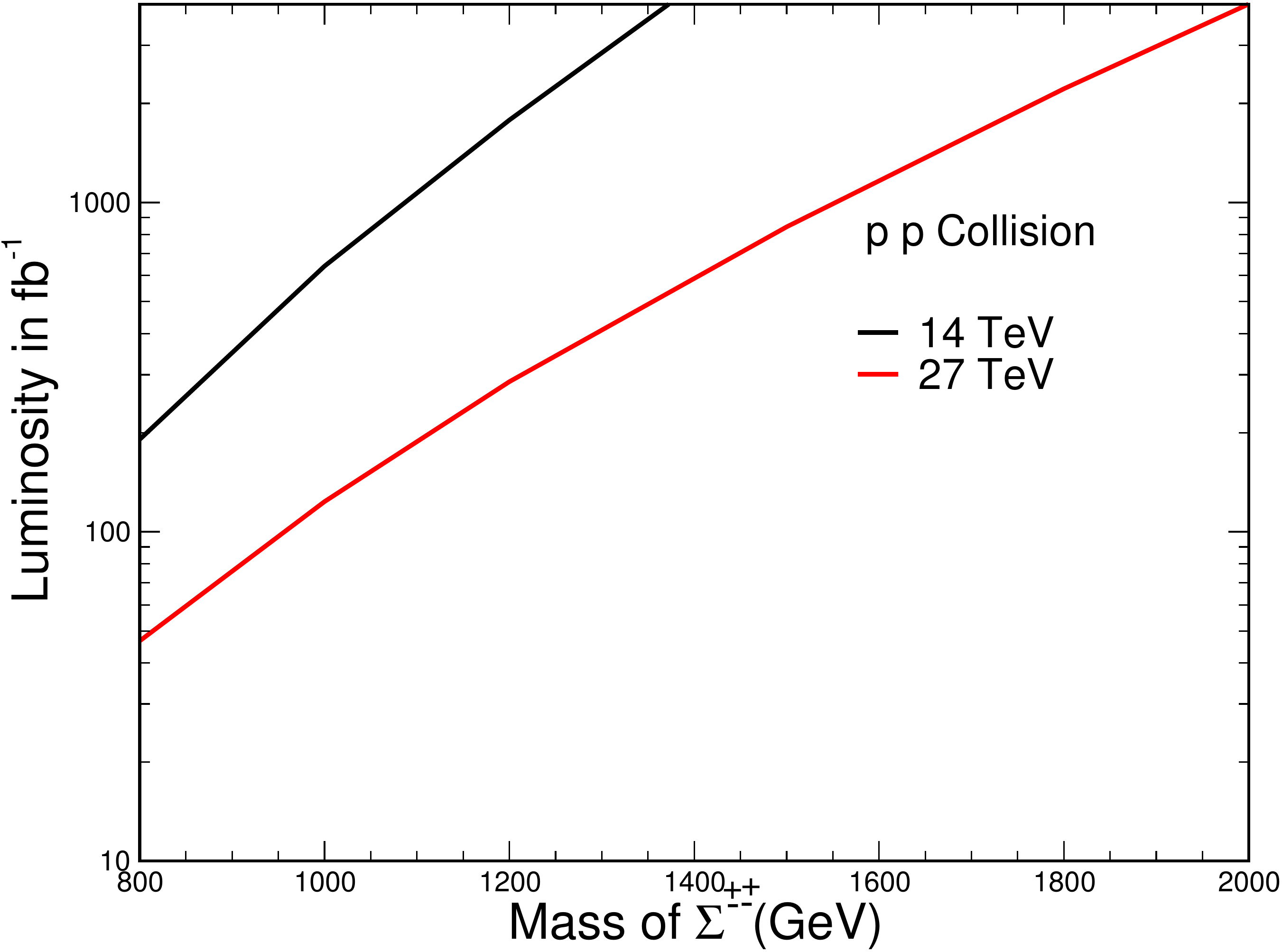}
\caption{The required luminosity to observed at least 10 events in 4 b-jets and at least one lepton final state 
for $p p \to Z/\gamma \to \Sigma^{++} \Sigma^{--}$ at 14 TeV and 27 TeV as a function of $\Sigma^{\pm\pm}$ mass.}
\label{fig:4}
\end{center}
\end{figure}

Finally we select the events that satisfy,\\
\begin{center}
$(M_{H}-15) $GeV$ < M_{2b} < (M_{H}+15)$ GeV\\
$(M_{\varphi_R}-200)$ GeV $< M_{4b} < (M_{\varphi_R}+200)$ GeV,
\end{center}
where $M_H$ is the SM Higgs mass.
After all the selections as mentioned above, the number of events to expect 
at 14 TeV LHC at different luminosities are given in Table~\ref{tab:2}. Note that,
we do not include the effect of the running of the coupling constant (see previous section),
as the mass of the scalar ($\varphi_R$)is less than 1 TeV, but for higher masses the running 
will be important. The number of events is given for both (4b-jets+ 1$l$) and (4-bjets+$l^+l^-$)
channels.
\begin{table}[!h]
  \begin{center}
    \caption{\label{tab:2} {\it Number of expected events at 150 $fb^{-1}$, 300 $fb^{-1}$ and 3000$fb^{-1}$ at 14 TeV p-p collision in different channels.}}
\begin{tabular}{|c|c|c|c|c|c|}
\hline
$\geq$ 4 bjets $\geq$ 1 ($l$) & $M_{\varphi_R}$ (GeV) & $\sigma$(fb) & N (150 $fb^{-1}$) & 300 $fb^{-1}$ & 3000 $fb^{-1}$ \\
\hline
&600&5.1&4.9&9.7&97\\
&700&2.5&2.3&4.6&45\\
&800&1.6&1.4&2.8&28\\
\hline
$\geq$ 4 bjets + 1 ($l^+l^-$) &&&&&\\
\hline
&600&5.1&4&8&81\\
&700&2.5&2&4&40\\
&800&1.6&1.2&2.4&24\\
\hline										
\hline
\end{tabular}
\end{center}
\end{table}
The number of events in each channel will further improve at a higher center of 
mass energy, 27 TeV. Hence the luminosity reach as a function of $\varphi_R$ mass is given 
in Figure~\ref{fig:4} for 14 TeV and 27 TeV center of mass energy at LHC.

\section{ Conclusions and discussions}
 In this paper, we have considered an extension of the SM, introducing large $SU(2)_L$ multiplet fields such as quartet, quintet and septet scalar fields, and Majorana quintet fermions. 
In our scenario, the quintet and septet scalars have vacuum expectation values which are constrained by the $\rho$-parameter, 
while the quartet scalar filed does not develop a VEV which is realized by assuming relation among the parameters in the potential.
Then, the active neutrino masses can be induced by interactions among these multiplets and the neutrinos at one loop level.
We have found that the neutrino masses are suppressed by the small VEVs of the septet and a loop factor,
explaining the smallness of the neutrino mass with relaxing the Yukawa hierarchies.
Carrying out numerical analysis, we find the neutrino data can be accommodated with $\mathcal{O}(0.1)$ to $\mathcal{O}(1)$ Yukawa couplings 
taking extra particle masses at TeV scale.

We have also discussed the collider physics considering production processes of charged particles in the large multiplets
especially focusing on doubly charged fermion from quintet.
The doubly charged scalar decays into lepton and components of quartet scalar via Yukawa interaction generating neutrino mass.
The components of quartet scalar decay via gauge interaction and/or interactions in scalar potential.
We then obtain signal of multi Higgs boson plus charged leptons and/or jets with/without missing transverse energy.
It has been shown that we can test our model in future LHC experiments estimating number of events imposing specific  kinematical cuts. We have shown the discovery potential of this background free channel at current and future luminosities, at LHC.


\section*{Acknowledgments}
\vspace{0.5cm}
This research was supported by an appointment to the JRG Program at the APCTP through the Science and Technology Promotion Fund and Lottery Fund of the Korean Government. This was also supported by the Korean Local Governments - Gyeongsangbuk-do Province and Pohang City (H.O.). H. O. is sincerely grateful for the KIAS member.
N.K. acknowledges 
the support from the Dr. D. S. Kothari Postdoctoral scheme (201819-PH/18-19/0013).
N. K. also acknowledge ``(9/27-28 @APCTP HQ) APCTP Mini-Workshop 
- Recent topics on dark matter, neutrino, and their related phenomenologies" where the 
problem had been proposed an also thanks the hospitality of APCTP, Korea.

\appendix
\section{ Appendix: $SU(2)_L$ large multiplet fields}

In this appendix we summarize expression of quartet scalar and quintet fermion.

\noindent
{\bf Scalar quartet field}

The quartet $\Phi_4$ with hypercharge $Y=1/2$ can be written as 
\begin{equation}
\Phi_4 = \left( \varphi^{++}, \varphi^{+}_2, \varphi^{0}, \varphi^{-}_1 \right)^T, \quad {\rm or} \quad
(\Phi_4)_{ijk}, 
\end{equation}
where $(\Phi_4)_{ijk}$ is the symmetric tensor notation denoted by 
$(\Phi_4)_{[111]} = \varphi^{++}$, $(\Phi_4)_{[112]} = \varphi^{+}_2/\sqrt{3}$, $(\Phi_7)_{[122]} = \varphi^{0}/\sqrt{3}$ and $(\Phi_4)_{[222]} = \varphi^{-}_1$; $[ijk]$ indices are symmetric under exchange among them.
Using the expression, we obtain
\begin{align}
\Phi_4^\dagger \Phi_4 &=   (\Phi_4^*)_{ijk} (\Phi_4)_{ijk} \nonumber \\
&= \varphi^{++} \varphi^{--} + \varphi^{+}_1 \varphi^{-}_1  + \varphi^{+}_2 \varphi^{-}_2  +  \varphi^0 \varphi^{0} 
\end{align}
where the iterated indices are always summed out. 
Then covariant derivative of $\Phi_4$ is given by
\begin{equation}
D^\mu \Phi_4 = \partial^\mu \Phi_4 - i \left( g W_a^\mu {\cal T}_a^{(4)} +  \frac{1}{2} g' B^\mu \right) \Phi_4,
\end{equation}
where $g(g')$ is the $SU(2)_L(U(1)_Y)$ gauge coupling and ${\cal T}^{(4)}_a$ denotes matrices for the generators of SU(2) acting on $\Phi_4^{}$ such that
 \begin{align}
{\cal T}^1 = \frac{1}{2}  \left( \begin{array}{cccc}
   0 & \sqrt{3} & 0 & 0  \\ 
    \sqrt{3} & 0 & 2 & 0  \\ 
    0 & 2 & 0 & \sqrt{3}  \\ 
    0 & 0 & \sqrt{3} & 0 \\ 
  \end{array}  \right)\,, \ \ \ {\cal T}^2 = \frac{i}{2}  \left( \begin{array}{cccc}
   0 & -\sqrt{3} & 0 & 0  \\ 
    \sqrt{3} & 0 & -2 & 0 \\ 
    0 & 2 & 0 & -\sqrt{3} \\ 
    0 & 0 & \sqrt{3} & 0  \\ 
  \end{array}  \right)\,,
 \end{align}
and ${\cal T}^3 = {\rm diag}(3/2, 1/2,  -1/2, -3/2)$.
The covariant derivative in terms of mass eigenstate of SM gauge boson can be derived applying $W^\pm_\mu = (W_{1 \mu} \mp W_{2 \mu})/\sqrt{2}$, $Z_\mu = \cos \theta_W W_{3 \mu} - \sin \theta_W B_\mu$ and $A_\mu = \sin \theta_W W_{3 \mu} + \cos \theta_W B_\mu$ where $\theta_W$ is the Weinberg angle.  Then we obtain the covariant derivative in terms of mass eigenstates of gauge bosons as follows
\begin{align}
(D_\mu \Phi_4)_m =& \left[ \partial_\mu - i \left(\frac12+m \right) e A_\mu - i \frac{g}{c_W} \left(m - \left( \frac12+m \right) s_W^2 \right) Z_\mu \right] (\Phi_4)_{m} \nonumber \\
& + \frac{i}{\sqrt{2}} \sqrt{ \left(\frac32 + m \right) \left(\frac52 -m \right) } W^+_\mu (\Phi_4)_{m-1}  + \frac{i}{\sqrt{2}} \sqrt{ \left(\frac32 - m \right) \left(\frac52 +m \right) } W^-_\mu (\Phi_4)_{m+1} ,
\end{align}
where the subscript $m$ distinguish component of the multiplet in terms of the eigenvalue of ${\cal T}^3$. \\

\noindent
{\bf Fermion quintet field}

The fermion quintet $\Sigma_R$ with hypercharge $Y=0$ can be written by 
\begin{equation}
\Sigma = \left[ \Sigma_1^{++}, \Sigma^{+}_1, \Sigma^{0}, \Sigma^{-}_2, \Sigma_2^{--} \right]_R^T, \quad {\rm or} \quad
(\Sigma_R)_{ ijkl}, 
\label{eq:sigmaRapp}
\end{equation}
where $(\Sigma_R)_{ijkl}$ is the symmetric tensor notation given by
$(\Sigma_R)_{[1111]} = \Sigma_{1R}^{++}$, $(\Sigma_4)_{[1112]} = \Sigma_{1R}^{+}/2$, $(\Sigma_R)_{[1122]} = \Sigma^{0}_R/\sqrt{6}$, $(\Sigma_R)_{[1222]} = -\Sigma^{-}_{2R}/2$ and $(\Sigma_R)_{[2222]} = \Sigma^{--}_{2R}$.
Using the expression, we obtain 
\begin{align}
\bar \Sigma_R^c \Sigma_R =&  (\bar \Sigma_R^c)_{ijkl} (\Sigma_R)_{i'j'k'l'} \epsilon^{ii'} \epsilon^{jj'} \epsilon^{kk'} \epsilon^{ll'}   \nonumber \\
= & \bar \Sigma^{++c}_{1R} \Sigma^{--}_{2R} + \bar \Sigma^{+c}_{1R} \Sigma^{-}_{2R} + \bar \Sigma^{0c}_{R} \Sigma^{0}_{R} + \bar \Sigma^{-c}_{2R} \Sigma^{+}_{1R} + \bar \Sigma^{--c}_{2R} \Sigma^{++}_{1R},
\end{align}
where $\epsilon^{ij}$ is anti-symmetric tensor.
The covariant derivative of $\Sigma_R$ can be derived as 
\begin{equation}
D^\mu \Sigma_R = \partial^\mu \Sigma_R - i g W_a^\mu {\cal T}_a^{(5)} \Sigma_R,
\end{equation}
where ${\cal T}^{(5)}_a$ denote the matrices for the generators of SU(2) acting on $\Sigma_R$ given by
\begin{eqnarray} & \displaystyle
{\cal T}_1^{(5)} \,\,=\,\, \frac{1}{2}
\begin{pmatrix}
0 & 2 & 0 & 0 & 0 \\
2 & 0 & \sqrt{6} & 0 & 0 \\
0 & \sqrt{6} & 0 & \sqrt{6} & 0 \\
0 & 0 & \sqrt{6} & 0 & 2 \\
0 & 0 & 0 & 2 & 0
\end{pmatrix} , \hspace{5ex}
{\cal T}_2^{(5)} \,\,=\,\, \frac{i}{2}
\begin{pmatrix}
0 & -2 & 0 & 0 & 0 \\
2 & 0 & -\sqrt{6} & 0 & 0 \\
0 & \sqrt{6} & 0 & -\sqrt{6} & 0 \\
0 & 0 & \sqrt{6} & 0 & -2 \\
0 & 0 & 0 & 2 & 0
\end{pmatrix} ,
& \nonumber \\ & \displaystyle
{\cal T}_3^{(5)} \,\,=\,\, {\rm diag}(2,1,0,-1,-2) ~.
\end{eqnarray}
The covariant derivative in terms of mass eigenstates of gauge bosons is derived as 
\begin{align}
(D_\mu \Sigma_R)_m = &   \left( \partial_\mu - i m e A_\mu -i g c_W m Z_\mu  \right) (\Sigma_R)_m \nonumber \\
&  + \frac{ig}{\sqrt{2}} \sqrt{(2+m)(3-m)} W_\mu^+  (\Sigma_R)_{m-1} + \frac{ig}{\sqrt{2}} \sqrt{(2-m)(3+m)} W_\mu^-  (\Sigma_R)_{m+1} .
\end{align}


\begin{thebibliography}{99}


\bibitem{Cirelli:2005uq} 
  M.~Cirelli, N.~Fornengo and A.~Strumia,
  Nucl.\ Phys.\ B {\bf 753}, 178 (2006)
  [hep-ph/0512090].
  
\bibitem{Hambye:2009pw} 
  T.~Hambye, F.-S.~Ling, L.~Lopez Honorez and J.~Rocher,
  JHEP {\bf 0907}, 090 (2009)
  Erratum: [JHEP {\bf 1005}, 066 (2010)]
  [arXiv:0903.4010 [hep-ph]].
  
\bibitem{delAguila:2013yaa} 
  F.~del Aguila, M.~Chala, A.~Santamaria and J.~Wudka,
  Phys.\ Lett.\ B {\bf 725}, 310 (2013)
  [arXiv:1305.3904 [hep-ph]].
  
\bibitem{delAguila:2013mia} 
  F.~del \'Aguila and M.~Chala,
  JHEP {\bf 1403}, 027 (2014)
  [arXiv:1311.1510 [hep-ph]].





\bibitem{Alvarado:2014jva} 
  C.~Alvarado, L.~Lehman and B.~Ostdiek,
  JHEP {\bf 1405}, 150 (2014)
  [arXiv:1404.3208 [hep-ph]].
  
\bibitem{Geng:2014oea} 
  C.~Q.~Geng, L.~H.~Tsai and Y.~Yu,
  Phys.\ Rev.\ D {\bf 91}, no. 7, 073014 (2015)
  [arXiv:1411.6344 [hep-ph]].
    
  
\bibitem{Aranda:2015xoa} 
  A.~Aranda and E.~Peinado,
  Phys.\ Lett.\ B {\bf 754}, 11 (2016)
  [arXiv:1508.01200 [hep-ph]].
    
\bibitem{Sierra:2016qfa} 
  D.~Aristizabal Sierra, C.~Simoes and D.~Wegman,
  JHEP {\bf 1606}, 108 (2016)
  [arXiv:1603.04723 [hep-ph]].

  
\bibitem{Sierra:2016rcz} 
  D.~Aristizabal Sierra, C.~Simoes and D.~Wegman,
  JHEP {\bf 1607}, 124 (2016)
  [arXiv:1605.08267 [hep-ph]].
  
  
  
\bibitem{Nomura:2016jnl} 
  T.~Nomura, H.~Okada and Y.~Orikasa,
  Phys.\ Rev.\ D {\bf 94}, no. 5, 055012 (2016)
  [arXiv:1605.02601 [hep-ph]].
  
\bibitem{Harris:2017ecz} 
  M.~J.~Harris and H.~E.~Logan,
  Phys.\ Rev.\ D {\bf 95}, no. 9, 095003 (2017)
  [arXiv:1703.03832 [hep-ph]].


\bibitem{Nomura:2017abu} 
  T.~Nomura and H.~Okada,
  Phys.\ Rev.\ D {\bf 96}, no. 9, 095017 (2017)
  [arXiv:1708.03204 [hep-ph]].



\bibitem{Chala:2018ari} 
  M.~Chala, C.~Krause and G.~Nardini,
  arXiv:1802.02168 [hep-ph].

\bibitem{Nomura:2018ktz} 
  T.~Nomura and H.~Okada,
  Phys.\ Lett.\ B {\bf 792}, 424 (2019)
  [arXiv:1809.06039 [hep-ph]].
  
\bibitem{Nomura:2018ibs} 
  T.~Nomura and H.~Okada,
  Phys.\ Rev.\ D {\bf 99}, no. 5, 055033 (2019)
  [arXiv:1806.07182 [hep-ph]].
  
\bibitem{Nomura:2018lsx} 
  T.~Nomura and H.~Okada,
  Phys.\ Dark Univ.\  {\bf 26}, 100359 (2019)
  [arXiv:1808.05476 [hep-ph]].
    
\bibitem{Nomura:2018cle} 
  T.~Nomura and H.~Okada,
  Phys.\ Lett.\ B {\bf 783}, 381 (2018)
  [arXiv:1805.03942 [hep-ph]].
  
\bibitem{Nomura:2018cfu} 
  T.~Nomura and H.~Okada,
  Phys.\ Rev.\ D {\bf 99}, no. 5, 055027 (2019)
  [arXiv:1807.04555 [hep-ph]].







\bibitem{Agashe:2014kda} 
  K.~A.~Olive {\it et al.} [Particle Data Group],
  Chin.\ Phys.\ C {\bf 38}, 090001 (2014).
  
  


\bibitem{Maki:1962mu} 
  Z.~Maki, M.~Nakagawa and S.~Sakata,
  Prog.\ Theor.\ Phys.\  {\bf 28}, 870 (1962).
  
\bibitem{Casas:2001sr} 
  J.~A.~Casas and A.~Ibarra,
  Nucl.\ Phys.\ B {\bf 618}, 171 (2001)
  [hep-ph/0103065].
  
  
  
\bibitem{Aubert:2009ag} 
  B.~Aubert {\it et al.} [BaBar Collaboration],
  Phys.\ Rev.\ Lett.\  {\bf 104}, 021802 (2010)
  [arXiv:0908.2381 [hep-ex]].
  
\bibitem{TheMEG:2016wtm} 
  A.~M.~Baldini {\it et al.} [MEG Collaboration],
  Eur.\ Phys.\ J.\ C {\bf 76}, no. 8, 434 (2016)
  [arXiv:1605.05081 [hep-ex]].

\bibitem{Renga:2018fpd} 
  F.~Renga [MEG Collaboration],
  Hyperfine Interact.\  {\bf 239}, no. 1, 58 (2018)
  [arXiv:1811.05921 [hep-ex]].
  
  
\bibitem{Hagiwara:2011af} 
  K.~Hagiwara, R.~Liao, A.~D.~Martin, D.~Nomura and T.~Teubner,
  J.\ Phys.\ G {\bf 38}, 085003 (2011)
  doi:10.1088/0954-3899/38/8/085003
  [arXiv:1105.3149 [hep-ph]].
  
  
\bibitem{Kanemura:2015bli} 
  S.~Kanemura, K.~Nishiwaki, H.~Okada, Y.~Orikasa, S.~C.~Park and R.~Watanabe,
  PTEP {\bf 2016}, no. 12, 123B04 (2016)
  [arXiv:1512.09048 [hep-ph]].
  
  

\bibitem{Alloul:2013bka}
A.~Alloul, N.~D. Christensen, C.~Degrande, C.~Duhr, and B.~Fuks, {\it
  {FeynRules 2.0 - A complete toolbox for tree-level phenomenology}},  {\em
  Comput. Phys. Commun.} {\bf 185} (2014) 2250--2300,
  [\href{http://arxiv.org/abs/1310.1921}{{\tt arXiv:1310.1921}}].
\bibitem{Christensen:2008py}
N.~D. Christensen and C.~Duhr, {\it {FeynRules - Feynman rules made easy}},
  {\em Comput. Phys. Commun.} {\bf 180} (2009) 1614--1641,
  [\href{http://arxiv.org/abs/0806.4194}{{\tt arXiv:0806.4194}}].
\bibitem{Alwall:2014hca}
J.~Alwall, R.~Frederix, S.~Frixione, V.~Hirschi, F.~Maltoni, O.~Mattelaer,
  H.~S. Shao, T.~Stelzer, P.~Torrielli, and M.~Zaro, {\it {The automated
  computation of tree-level and next-to-leading order differential cross
  sections, and their matching to parton shower simulations}},  {\em JHEP} {\bf
  07} (2014) 079, [\href{http://arxiv.org/abs/1405.0301}{{\tt
  arXiv:1405.0301}}].
\bibitem{Ball:2012cx}
R.~D. Ball et~al., {\it {Parton distributions with LHC data}},  {\em Nucl.
  Phys.} {\bf B867} (2013) 244--289,
  [\href{http://arxiv.org/abs/1207.1303}{{\tt arXiv:1207.1303}}].
\bibitem{Ball:2014uwa}
{\bf NNPDF} Collaboration, R.~D. Ball et~al., {\it {Parton distributions for
  the LHC Run II}},  {\em JHEP} {\bf 04} (2015) 040,
  [\href{http://arxiv.org/abs/1410.8849}{{\tt arXiv:1410.8849}}].
\bibitem{Ball:2013hta}
{\bf NNPDF} Collaboration, R.~D. Ball, V.~Bertone, S.~Carrazza, L.~Del~Debbio,
  S.~Forte, A.~Guffanti, N.~P. Hartland, and J.~Rojo, {\it {Parton
  distributions with QED corrections}},  {\em Nucl. Phys.} {\bf B877} (2013)
  290--320, [\href{http://arxiv.org/abs/1308.0598}{{\tt arXiv:1308.0598}}].
\bibitem{Martin:2004dh}
A.~D. Martin, R.~G. Roberts, W.~J. Stirling, and R.~S. Thorne, {\it {Parton
  distributions incorporating QED contributions}},  {\em Eur. Phys. J.} {\bf
  C39} (2005) 155--161, [\href{http://arxiv.org/abs/hep-ph/0411040}{{\tt
  hep-ph/0411040}}].


\bibitem{Schmidt:2015zda}
C.~Schmidt, J.~Pumplin, D.~Stump, and C.~P. Yuan, {\it {CT14QED parton
  distribution functions from isolated photon production in deep inelastic
  scattering}},  {\em Phys. Rev.} {\bf D93} (2016), no.~11 114015,
  [\href{http://arxiv.org/abs/1509.02905}{{\tt arXiv:1509.02905}}].

\bibitem{Aaboud:2018knk} 
  M.~Aaboud {\it et al.} [ATLAS Collaboration],
  JHEP {\bf 1901}, 030 (2019)
  doi:10.1007/JHEP01(2019)030
  [arXiv:1804.06174 [hep-ex]].
  
\end{thebibliography}
\end{document}